\documentclass[11pt]{article}
\usepackage[utf8]{inputenc}
\usepackage[T1]{fontenc}
\usepackage{graphicx}
\usepackage{longtable}
\usepackage{wrapfig}
\usepackage{rotating}
\usepackage[normalem]{ulem}
\usepackage{amsmath}
\usepackage{amssymb}
\usepackage{capt-of}
\usepackage{hyperref}
\usepackage{float}
\usepackage{booktabs}
\usepackage{array}
\usepackage[margin=1in]{geometry}
\usepackage{placeins}
\usepackage[numbers]{natbib}
\hypersetup{hypertexnames=false}

\date{}
\title{Federated Sharing and Continuous Improvement of Medical Device Knowledge Artifacts: A Conceptual Model}
\hypersetup{
 pdfauthor={J.C. Mariscal-Melgar, Sonja Buxbaum-Conradi, Victoria Wenzelmann, Tobias Redlich},
 pdftitle={Federated Sharing and Continuous Improvement of Medical Device Knowledge Artifacts: A Conceptual Model},
 pdfkeywords={},
 pdfsubject={},
 pdfcreator={Emacs 30.0.50 (Org mode 9.6.10)}, 
 pdflang={English}}
\usepackage{natbib}
\begin{document}

\title{Federated Sharing and Continuous Improvement of Medical Device Knowledge Artifacts: A Conceptual Model}

\author{
J.C. Mariscal-Melgar, Sonja Buxbaum-Conradi, \\
Victoria Wenzelmann, and Tobias Redlich \\
\small Helmut-Schmidt University/University of the Federal Armed Forces Hamburg \\
\small Corresponding author: \texttt{jcm@hsu-hh.de}
}

\date{}
\maketitle

\begin{abstract}
Healthcare organisations use digital systems to exchange information from
clinical cases. Medical centres with digital production facilities create device
designs during care. These designs and production records often remain at the
site that made them. Other sites may struggle to find a suitable design or learn
what happened when staff used it. Mobile medical centres may also lose access
when they work away from hospital systems. This paper proposes an
artifact-centred model for a federated exchange infrastructure that lets
hospitals and mobile medical centres share and improve medical knowledge while
controlling their own records and decisions. An integrative literature review
screened 910 records and mapped 240 publications across six questions. We read
72 publications in detail to trace the path from local use to a decision about
shared knowledge. The review found no common process that links a record of
local use to a decision about changing the knowledge shared with later users.
The model keeps case data at each site and records which artifact version
informed each use. Federation lets sites share reviewed versions and return
records from use for review. Mobile units can receive artifacts before
deployment and record their use while offline. After reconnecting, they can
exchange these records with other sites.  Point-of-care manufacturing shows how
a medical device knowledge artifact connects a design to production records
while the care site controls product release. Federation could form a governed
learning network where one site's experience improves medical knowledge
artifacts used elsewhere while authority remains local.
\end{abstract}


\section{Introduction}
\label{sec:org00bd236}

Medical organisations collect large amounts of information, but only some of it
is turned into knowledge that improves over time. For example, a clinical
protocol may sit in a document store while its research is held elsewhere. An
engineering design may remain in a technical database even after the case that
motivated it has closed. As a result, users can find the content without seeing
where it applies or what happened when others used it. Simply providing access
to separate pieces of information does not create a cumulative learning process
\citep{FriedmanEtAl2016LHS,AlperEtAl2021Metadata}.

Point-of-care manufacturing makes this separation visible. For the workflow that
motivates this paper, no single clinician-facing record links the clinical need
to the medical device knowledge artifact used for production. The missing links
include the computer-aided design (CAD) version and the records made during
fabrication. Planning and fabrication use different tools. Inspection and
release create further records.  Staff can document each activity, but they lack
one path that preserves the design history and returns production experience to
the knowledge used in the next case.  Studies of hospital additive manufacturing
report the same need to connect local capability with quality controls and
assigned responsibility
\citep{ChristensenRybicki2017Printing,VallsEsteveEtAl2024PoC}.

Reuse alone does not solve this problem. A team reuses knowledge when it applies
an existing artifact to a new task. Continuous improvement requires the team to
record which version it selected and whether it changed that version. The record
must also explain the conditions of use and the observed result. A reviewer can
then decide whether the evidence supports a correction or a successor. Evidence
that applies only in one setting may instead support a variant. Without this
decision, repeated reuse produces more copies rather than better shared
knowledge.

Federated infrastructures already show how organisations can cooperate without
placing all source data in one repository. The Sentinel programme distributes
analyses to data partners \citep{PlattEtAl2012MiniSentinel}. The Observational
Medical Outcomes Partnership common data model and the Observational Health Data
Sciences and Informatics network support common analyses across separate
databases \citep{VossEtAl2015OMOP,HripcsakEtAl2016OHDSI}. Federated learning
applies a related principle to model development
\citep{RiekeEtAl2020Federated}. These approaches establish federation as prior
art and supply the network layer. The remaining design question is how a
reusable medical knowledge artifact changes after its application.

The unresolved question therefore concerns what an exchange infrastructure
should govern and how that object should change. This paper calls the common
object a \emph{medical knowledge artifact}. It is a governed, versioned unit of
reusable medical knowledge that states its purpose and scope. It also preserves
its change history and review status. A \emph{medical device knowledge artifact} is a medical
knowledge artifact that represents a reusable device design or production
method. For a CAD-based device, it preserves the editable design's history and
links each use to the version selected. It is neither the patient record nor the
physical item. These requirements arise within one organisation before exchange
across sites adds further governance.

This paper derives an artifact-centred model from a structured integrative
review. It makes five contributions. First, it defines a medical knowledge
artifact and a device-focused subtype, a medical device knowledge artifact, and
separates both from confidential case records. Second, it specifies a lifecycle
that connects local use to governed revision. Third, it derives requirements for
an exchange infrastructure that could support that lifecycle. Fourth, it shows
how federation can extend the lifecycle. Finally, it presents propositions and a
research agenda for future empirical studies. A point-of-care manufacturing
scenario is used to test the internal logic of the concept and to define
requirements for a first implementation study.

\section{Motivation}
\label{sec:org8370865}

\begin{figure}[H]
\centering
\includegraphics[width=0.96\linewidth]{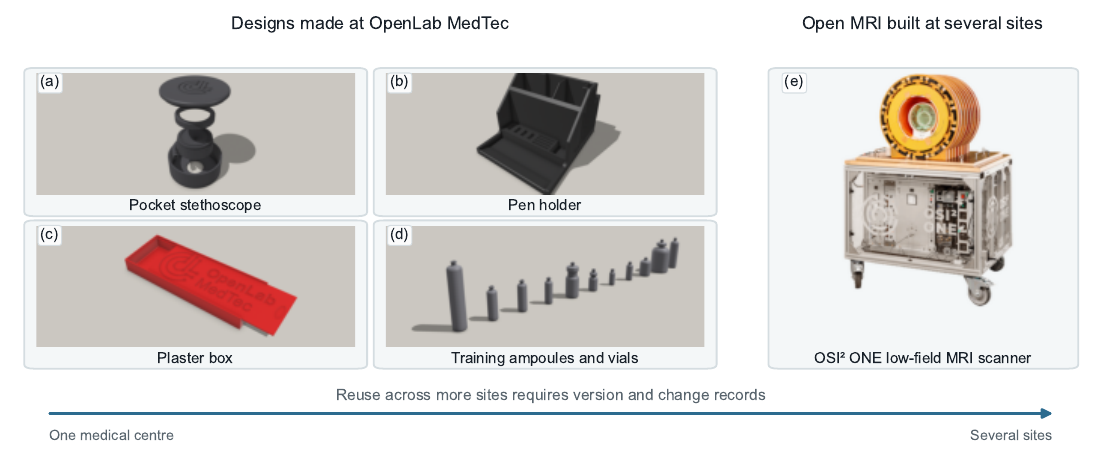}
\caption{\label{fig:org9abe521}From local design work to reuse across sites. Panels (a)--(d) show the four illustrative projects in the OpenLab MedTec. Panel (e) shows OSI² ONE, an open low-field magnetic resonance imaging scanner built at several sites.}
\end{figure}

Healthcare organisations sometimes need hardware that is unavailable or does
not fit the work at their site. Some hospitals respond by designing and making
hardware where staff provide care. When the result is a medical device, this
activity is called point-of-care manufacturing. The work produces a design.
Staff also learn how to make the item. Another medical centre may face a similar
need, but it cannot use the design if the file remains inside the first site. If
the original team does not record what it learned, staff elsewhere cannot use
that knowledge. OpenLab MedTec provides an example. The lab operates at the
Bundeswehr Hospital Hamburg and brings clinical staff together with technicians
who can design and make hardware. It provides fabrication equipment and design
support
\citep{OpenLabMedTec,NewProductionInstituteFabCity}. Its repository stores the
source and production files created during this work
\citep{OpenLabMedTecHardwareProjects}. The four projects in Figure
\ref{fig:org9abe521} show how designs can accumulate at one site.

Open hardware projects show why sharing a design becomes harder as the system
grows. The OpenFlexure Microscope combines a printed mechanism with a
software-controlled imaging system. Teams can build it locally and change its
optics for their work \citep{CollinsEtAl2020OpenFlexure}. OSI² ONE applies open
development to a modular low-field magnetic resonance imaging scanner
\citep{WinterEtAl2024OpenMRI,OpenSourceImagingOSIIone}. Teams have built
scanners in Leiden and Berlin. A third scanner was built in Mbarara
\citep{OpenSourceImagingOSIIone}. A builder must know which design it followed
and what changed during construction. Later builders also need to know which
changes passed review. Figure \ref{fig:org9abe521} shows objects made
inside one hospital beside the open-mri scanner reproduced at several
sites. This change in scale makes design identity and change records matter.

Published open-source hardware is prepared for people outside the team that
created it. A catalogue can state a project's purpose and source. It can also
state the manufacturing process and licence
\citep{TechnologyProductPortfolioOSH}. Designs made within a medical centre
often start differently. Staff may create a design for one case or a need at
their site, then save it under a local name. Another centre cannot search for a
design it does not know exists. If staff obtain the file by chance, its local
name may not explain whether the design fits their site. The discovery problem
is therefore not a shortage of designs. One centre searches by its need, while
another stores the answer as a local file.

Local fabrication can change a medical centre from a user of products into a
source of designs for other sites. That role raises questions that storing a
design file does not answer. A design file contains geometry, but another team
also needs to know why staff made each design choice and what happened during
production. The site that created the design can learn when another team uses
different equipment or changes the design for its setting. If this knowledge
stays in separate systems or with individuals, a centre may repeat work already
done elsewhere. The aim is to make local design work usable as shared knowledge.
This requires medical centres to learn from production and share that learning
across sites.

\section{Literature Review Methodology}
\label{sec:orgacbdd51}

Healthcare staff use medical knowledge when they follow a clinical protocol or
make a device from a design. This review compares work from six fields to ask
how staff record what happened and how authorised people decide whether that
knowledge should change.

\subsection{Review Aim and Design}
\label{sec:org21d6c24}

We used an integrative literature review because several fields study different
parts of this process. This type of review allows research with different
questions or methods to be examined together. The sources cover clinical work
and information systems. They also cover medical-device production. Unlike a
review of one treatment, this review does not combine outcome measurements. It
asks how staff record what happened when they used the knowledge and how that
record informs a later decision.

We read each source to determine which part of the process it explained. We
first identified the knowledge or file and who used it. We then asked how the
user obtained it and what they did with it. We looked for a record that named
the version used. When a source described a change, we recorded who proposed it
and who could approve it. Most sources explained only part of the process. We
therefore compared fields to see whether one field answered a question left open
by another. Six review questions organised the search and analysis (See
\emph{Interpretive Reading and Coding}). They determined how each publication was
grouped and compared.

\subsection{Selected Foundational Studies}
\label{sec:org492d3b3}

Before running the searches, we selected studies that described parts of the
process examined in this review. We chose studies that defined a learning
process or described a reusable object. Some also explained what organisations
may exchange while keeping source records at each site. These studies supplied
terms for the search queries. We call them \emph{selected foundational studies}
because they provided this starting point. The name does not rank them above
other studies. Citation counts helped us find related work, but they did not
decide which studies entered the set.

The selected foundational studies guided the search but did not enter the
review automatically. We screened each one under the same rules used for all
search results. Table \ref{tab:orgc971130} shows the search concept drawn
from each study and the question it helped form. This separation meant that a
study could help plan the search without automatically entering the final
collection of publications.

\begin{longtable}{@{}p{0.09\linewidth}p{0.23\linewidth}p{0.22\linewidth}p{0.31\linewidth}@{}}
\caption{\label{tab:orgc971130}Selected foundational studies used to orient the review.}
\\[0pt]
Year & Study & Search concept & Question carried into the review\\[0pt]
\hline
\endfirsthead
\multicolumn{4}{l}{Continued from previous page} \\[0pt]
\hline

Year & Study & Search concept & Question carried into the review \\[0pt]

\hline
\endhead
\hline\multicolumn{4}{r}{Continued on next page} \\
\endfoot
\endlastfoot
\hline
2003 & Peleg et al. \citep{Peleg2003GuidelineModels} & Computer-interpretable guideline & How does an executable knowledge object change after use?\\[0pt]
2007 & Institute of Medicine \citep{InstituteMedicine2007Learning} & Learning health system & What returns from practice to the knowledge used in later care?\\[0pt]
2010 & Brown et al. \citep{Brown2010Distributed} & Distributed health-data network & Which records remain with each organisation, and which may be exchanged?\\[0pt]
2013 & Ciccarese et al. \citep{Ciccarese2013PAV} & Authorship and version provenance & Which relation connects one version to another?\\[0pt]
2016 & Wilkinson et al. \citep{Wilkinson2016FAIR} & Reuse of digital resources & Which metadata make a resource findable and reusable?\\[0pt]
2016 & Friedman et al. \citep{FriedmanEtAl2016LHS} & Learning cycle & Which organisational steps connect care, analysis, and later action?\\[0pt]
2016 & Di Prima et al. \citep{DiPrimaEtAl2016FDA} & Additive manufacture of medical products & Which records connect a design to a production process?\\[0pt]
2018 & Flynn et al. \citep{FlynnEtAl2018KORO} & Computable biomedical knowledge object & What can an exchange infrastructure treat as a unit of reuse?\\[0pt]
2020 & Rieke et al. \citep{RiekeEtAl2020Federated} & Federated model training & What can sites compute together while source records remain local?\\[0pt]
2021 & Haro et al. \citep{CorpusHaro2021PointCareManufacturing} & Hospital production & Which duties arise when a hospital makes a patient-specific item?\\[0pt]
\end{longtable}

\subsection{Search Orientation and Query Design}
\label{sec:org810f94a}

Different fields use different words for the same process. We therefore studied
the terms used in each field before writing the database searches. Research on
hospital work led us to \emph{knowledge sharing} and \emph{knowledge management}. It also
introduced \emph{communities of practice}, a term for groups that learn through
shared work. Learning-health-system research supplied terms about feedback and
data reuse. Research on computable knowledge added \emph{repository} for storage and
\emph{provenance} for origin. It also added \emph{version control} for tracking change.

Federation allows organisations to work together while keeping source records
at each site. A \emph{distributed health-data network} links those sites, while a
\emph{common data model} gives them a shared structure for analysis. \emph{Federated
learning} lets sites contribute to a model without pooling their source
records. The search queries used this vocabulary and added terms for approval
across sites. The medical-device searches began with \emph{point-of-care
manufacturing}, in which a care organisation makes a device near the place of
care. We then followed the path from a medical image to a design file.
\emph{Segmentation} identifies the structures to model, while \emph{computer-aided
design} shapes the device. Searches for document control and open model
libraries tested whether researchers treated those files as objects that people
could share and manage over time.

A query string is the set of words submitted to a search service. We ran 25
query strings in OpenAlex, an index of scholarly publications. We also ran
focused searches in PubMed, a database of health and medical publications, to
check whether our terms found research about clinical use. The PubMed results
served only as a vocabulary check and did not enter the reported record counts.
We built the reproducible mapping corpus from the OpenAlex records. The search
script saves each query and every returned publication identifier. It also
saves the tags used during selection in the supplementary corpus files.

\subsection{Corpus Construction}
\label{sec:orga0045ad}

\begin{figure}[H]
\centering
\includegraphics[width=0.96\linewidth]{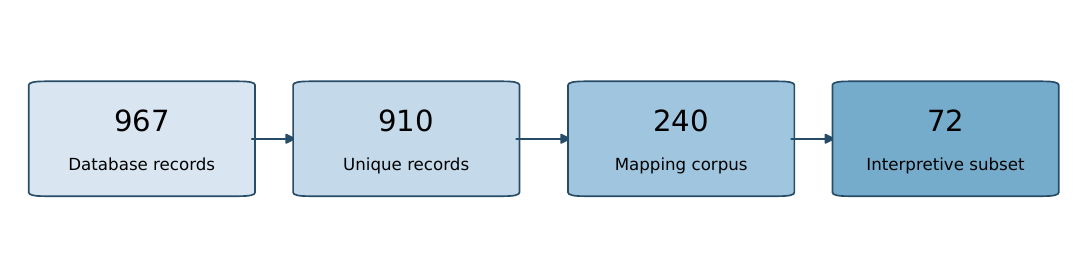}
\caption{\label{fig:org832d0ed}Construction of the mapping corpus and interpretive subset.}
\end{figure}

The database searches returned 967 records. The same publication could appear in
more than one search, so we removed duplicate entries. We first matched records
by digital object identifier, a code assigned to a publication. We then compared
titles. This process left 910 publications for screening. Figure \ref{fig:org832d0ed}
shows how this set was reduced.

We screened the title and abstract of each publication to decide whether it
could answer one review question. A study of knowledge sharing qualified only
if it examined people working in a clinical or hospital setting. A federation
study had to describe work coordinated across sites. This work could involve a
query sent to several sites or a model exchanged between them. It could instead
concern a shared way to represent data or a rule used across sites. A study of
hospital data qualified only if it explained the work needed to make local
records useful for another purpose.

Medical-device publications qualified only if they connected a design or
production method to medical use. The computer-aided design group included
segmentation studies, which examine how staff select structures from a medical
image. Studies of document control or model libraries also qualified when they
showed how design files were stored or shared. We excluded a publication when a
search term appeared but the study did not address a review question. We also
excluded three-dimensional models with no medical or related use. Production
studies were excluded unless they concerned a medical device or production
within a hospital setting.

After screening, the mapping corpus contained 240 publications. The \emph{mapping
corpus} is the collection used to compare publications assigned to the six
review questions. We selected 40 publications for each question. Giving each
question the same number prevented fields with more publications from
dominating the comparison. This balance supports comparison; it does not
estimate how much research each field contains.

We then selected 72 publications for closer reading, with 12 from each group.
We call this collection the \emph{interpretive subset} because it supplied the
detailed findings used in the synthesis. It retained every selected
foundational study that passed screening. We also chose publications that
explained how a process worked or what limited it. Some showed where the concept
might not apply. For the remaining places, we preferred journals or other
publication venues not yet represented in the same group.

\subsection{Quantitative Mapping}
\label{sec:orgfad1041}

Quantitative mapping describes the 240-publication corpus with counts rather
than interpretations. For each publication, we recorded its year and type. We
recorded where it was published and whether an open link to the full text was
available. We also noted search-group overlap, which means that more than one
search group returned the same publication. These values describe the chosen
corpus, not an entire field. A publication did not count as stronger support
merely because it had more citations.

The term-cloud map shows the words used in the titles and abstracts of each
review group. We removed common English words. We also removed words that only
describe sections or research procedures. We then calculated the
term-frequency--inverse-document-frequency for each remaining word. This measure
gives a word more weight when it occurs in one publication but is less common
across the corpus. We averaged these weights for the publications in each
group. A larger word in the figure therefore has a higher average weight in that
group. It does not mean that the word or publication is more important.

We divided publication years into four periods and counted how many
publications from each review group fell in each period. This map shows when the
selected publications for each question appeared. We did not use a topic model,
which is an automated method for finding groups of words in a collection. The
six groups had already been defined by the review questions and search process.
A topic model could therefore repeat the search vocabulary and make it appear
as a finding. Counts and dates provide a more direct description of the
publications used for close reading.

\subsection{Interpretive Reading and Coding}
\label{sec:orga86cb20}

Coding turns what publications report into information that can be compared.
We used the same set of labels when reading every publication. Each label is
called a \emph{code} and names one type of information to record. The review
question explained why a publication belonged in the corpus. The codes defined
what we recorded from it.

The review used six questions:

\begin{enumerate}
\item How do staff share knowledge in hospitals and clinics?
\item What work is needed before hospital data can serve as knowledge for another
task?
\item What do organisations exchange through federated health infrastructure, in
which each site keeps its source records?
\item How do people link one version of reusable medical knowledge to the next
when its content changes?
\item How do people preserve design history when they share medical
computer-aided design files?
\item How do hospitals manage the production of medical devices at the point of
care?
\end{enumerate}

Each publication was assigned to one main review question. A publication could
inform another question, but we counted it only in its main group. Each group
contained 40 publications. The six groups therefore formed a mapping corpus of
240 publications.

The first reading checked each group assignment. We read the title and abstract
of all 240 publications. A publication remained in its group only if it
addressed the assigned question.

We selected 72 publications for closer reading, with 12 from each group. This
subset retained the selected foundational studies that passed screening. It
also included publications that explained how a process worked or what limited
it. Some showed where the concept might not apply. We used the same codes for
every publication so that findings from the six groups could be compared.

During this reading, we traced what people did with the knowledge or file
described in each publication. We identified the item and its user. We then
recorded how the item reached the user and what the user did with it. If the
content changed, we recorded who proposed the change. We also recorded who had
authority to approve it.

Table \ref{tab:org0a9b367} lists the codes used for all 72 publications. In the
table, \emph{item} is a short name for the knowledge or file discussed by a
publication. \emph{User} means the person or organisation that relied on the item.
The last column states what we recorded under each code.

\begin{longtable}{@{}p{0.19\linewidth}p{0.36\linewidth}p{0.35\linewidth}@{}}
\caption{\label{tab:org0a9b367}Codes and information recorded during interpretive reading.}
\\[0pt]
Label & Meaning in this review & What was recorded\\[0pt]
\hline
\endfirsthead
\multicolumn{3}{l}{Continued from previous page} \\[0pt]
\hline

Label & Meaning in this review & What was recorded \\[0pt]

\hline
\endhead
\hline\multicolumn{3}{r}{Continued on next page} \\
\endfoot
\endlastfoot
\hline
Item & The knowledge or file examined by the publication & Its name and form\\[0pt]
User & A person or organisation that relied on the item & Who used it\\[0pt]
Sharing & Someone made the item available to another person or organisation & Who sent it and who received it\\[0pt]
Application & Someone relied on the item to perform a task & Who used it and what they did\\[0pt]
Where the item moved & The item passed between professions or organisations & Where it started and where it went\\[0pt]
Access & A person or organisation could obtain the item & How they obtained it\\[0pt]
Reuse & Someone applied an existing version in later work & Which version they used again\\[0pt]
Proposed change & Someone suggested a new version & Who proposed it and why\\[0pt]
File edit & The content of a file changed & What changed\\[0pt]
Review decision & A person with authority set the status of a proposed version & Who decided, what they decided, and the status\\[0pt]
Type of change & A correction fixed an error; a local adaptation served one setting & Which type of change the publication described\\[0pt]
\end{longtable}

The use codes distinguish finding an item from relying on it. \emph{Access} means
that a person or organisation could obtain the item. \emph{Application} means that
someone relied on it to perform a task. \emph{Reuse} means that someone applied an
existing version in later work. Finding a file did not by itself count as
application or reuse.

The change codes distinguish editing content from accepting a new version.
\emph{File edit} records what changed in the content. \emph{Proposed change} records who
requested a new version and why. \emph{Review decision} records what a person with
authority decided. This separation prevents every edit from being counted as
an improvement.

Artifact review remained separate from decisions about a patient or produced
item. In this analysis, a medical knowledge artifact could be a protocol or
design whose versions people manage and review. The codes recorded whether a
person with authority changed the status of that artifact. A patient-care
decision did not approve the artifact. Neither did a decision to release a
product. Regulatory conformity remained outside the coding process.

We read the full text when a title and abstract did not answer a coding
question. We also checked the full text before using a publication to support a
claim in the literature synthesis. This check connected each synthesized claim
to what its source reported.

\subsection{Thematic Synthesis}
\label{sec:org0c37646}

A \emph{theme} in this review is a relationship that several publications describe.
It is not a word that appears often in titles or abstracts. \emph{Thematic synthesis}
is the process used to find these relationships in the coded information and
connect findings from different review questions.

The first stage compared publications assigned to the same review question. We
identified the knowledge or file in each publication and how people used it. We
then examined how the item passed between people or organisations. If its
content changed, we recorded who had authority to approve the change. We treated
a relationship as a finding for the group only when several publications
described it.

The second stage compared findings from all six review groups. A relationship
entered the conceptual model only if publications from at least two groups
described it. This rule prevented the model from relying on the view of one
field.

The absence of an explanation in the publications did not prove that a practice
was absent. We used \emph{under-specified} when the corpus described parts of a
process but did not explain how those parts connected.

Computer-aided design history shows how we applied this rule. Some publications
described file controls within one site. Others described resources for sharing
designs. They did not explain how an adaptation made by one hospital could
enter a review process for use beyond that hospital. We therefore call this
connection under-specified. This does not mean that hospitals lack tools for
tracking file versions.

\section{Literature Review Synthesis}
\label{sec:orged611eb}

Medical knowledge can take the form of a clinical protocol or a design used to
make a device. This paper asks how an organisation can record what happened
when staff used that knowledge and decide whether it should change for later
users. We synthesize six review groups because none describes this whole
process. (1) The hospital knowledge-sharing group examines how staff make their
experience usable by others. (2) The group on hospital data examines the work
needed before records can support another task. (3) The federation group shows
how organisations exchange queries or models while keeping source records at
each site. (4) The group on medical knowledge artifacts examines how rules and
guidelines are stored and linked across versions. (5) The computer-aided design
group follows how design files are created and shared. (6) The point-of-care
production group examines how a hospital makes an item and decides whether to
release it. Together, the six groups explain parts of the process, but the
reviewed publications do not link the version used in local work to a later
decision about changing shared knowledge. We call the unit that makes this link
a \emph{medical knowledge artifact}. A protocol or device design can be managed as
such an artifact. Its record states its purpose and where it applies. It keeps
its change history and review status. Each use can then name the version that
informed the work, and a person with authority can decide whether the result
supports a change to shared knowledge. The term does not turn all clinical
knowledge into files. It applies only to knowledge that can be recorded in a
reusable form and placed under review. By showing what prior work provides and
what remains missing, the synthesis supplies the basis for the conceptual model
and its requirements.

\subsection{Corpus Map}
\label{sec:org6d63958}

The corpus map uses counts to describe the publications selected for this
review. The mapping corpus contains 240 publications dated from 1999 to
\begin{enumerate}
\item Of these, 223 are research articles and 11 are reviews. Four are book
\end{enumerate}
chapters. One is a report and one is a data paper. Table
\ref{tab:org94743f8} divides the publications among the six review groups.
Each group contributes 40 publications to the mapping corpus and 12 to the
subset used for closer reading. Equal group sizes allow comparison; they do not
mean that each field contains the same amount of research.

\begin{longtable}{@{}p{0.43\linewidth}p{0.14\linewidth}p{0.14\linewidth}p{0.18\linewidth}@{}}
\caption{\label{tab:org94743f8}Composition of the mapping corpus and interpretive subset.}
\\[0pt]
Review group & Mapping corpus & Interpretive subset & Date range\\[0pt]
\hline
\endfirsthead
\multicolumn{4}{l}{Continued from previous page} \\[0pt]
\hline

Review group & Mapping corpus & Interpretive subset & Date range \\[0pt]

\hline
\endhead
\hline\multicolumn{4}{r}{Continued on next page} \\
\endfoot
\endlastfoot
\hline
Clinical knowledge sharing and organisational learning & 40 & 12 & 2006--2024\\[0pt]
Hospital data and knowledge-building shortcomings & 40 & 12 & 2015--2026\\[0pt]
Federated hospital data and model infrastructures & 40 & 12 & 2010--2025\\[0pt]
Versioned medical knowledge artifacts & 40 & 12 & 1999--2026\\[0pt]
Medical computer-aided design history and sharing & 40 & 12 & 2008--2026\\[0pt]
Point-of-care medical-device production & 40 & 12 & 2015--2026\\[0pt]
\textbf{Total} & \textbf{240} & \textbf{72} & \textbf{1999--2026}\\[0pt]
\end{longtable}

\begin{figure}[H]
\centering
\includegraphics[width=0.96\linewidth]{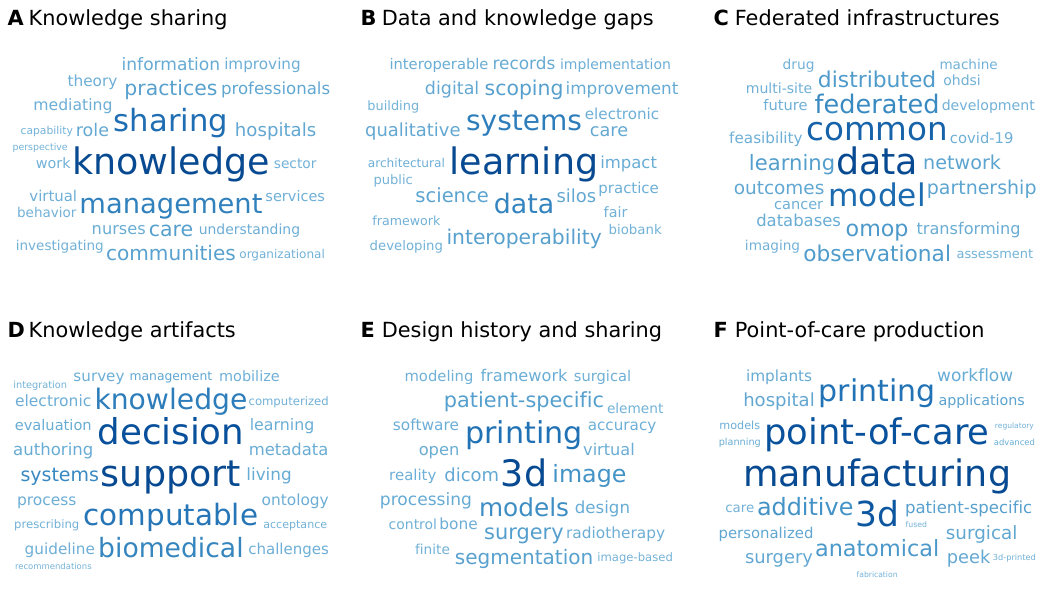}
\caption{\label{fig:org832fb37}Term clouds for the six review groups. Each panel uses the titles and abstracts of the 40 publications assigned to that group. Word size represents the group's mean term-frequency--inverse-document-frequency weight after common English words and paper-structure terms were removed. The review groups were set before this analysis; the clouds do not represent inferred topics.}
\end{figure}

The review groups use different terms for their subjects. Figure
\ref{fig:org832fb37} contains one term cloud for each group. A larger word
has a higher average weight in the titles and abstracts from that group. The
knowledge-sharing group gives more weight to \emph{knowledge} and \emph{sharing}, while
the federation group gives more weight to \emph{data} and \emph{models}. \emph{Printing}
appears in both device groups. \emph{Segmentation} carries more weight in the design
group, while \emph{manufacturing} carries more in the production group.

The clouds reveal a shift in what each group treats as the object of work. The
knowledge-sharing panel centres on people and work, while the data-gap panel
centres on systems and interoperability. The federation panel gives most weight
to data and models. The knowledge-artifact panel connects computable knowledge
with decision support. The two device panels follow images and designs into
printing or manufacturing. Across all six panels, terms that connect a use to a
version and then to review do not dominate the vocabulary. This pattern helps
locate the question examined by the paper, but it does not answer it. The search
and screening process defined the groups before the words were counted, so
visible terms can reflect those choices. Claims in the synthesis therefore
come from reading the publications and comparing coded relationships.

\begin{figure}[H]
\centering
\includegraphics[width=0.96\linewidth]{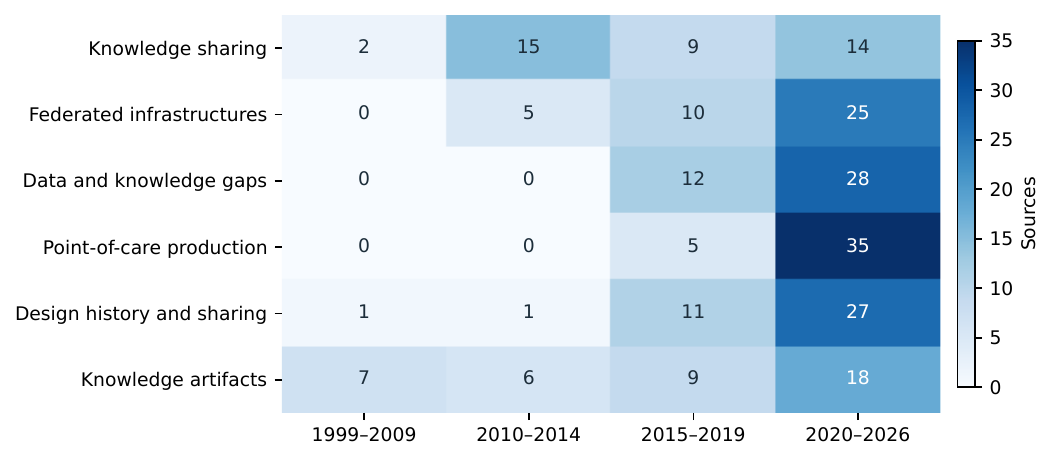}
\caption{\label{fig:org84cc270}Publication periods within each balanced review group. Counts describe the selected corpus, not the size of each field.}
\end{figure}

The date map shows that the six review groups did not appear at the same time
within the selected corpus. Knowledge sharing and medical knowledge artifacts
appear in every period. Their distributions differ: knowledge sharing peaks in
2010--2014, while publications on medical knowledge artifacts rise from 7
in 1999--2009 to 18 in 2020--2026. Federation grows from 5 publications in
2010--2014 to 25 in 2020--2026. Research on the gap between hospital data and
reusable knowledge enters the corpus in 2015--2019 and rises from 12 to 28
publications. The device groups show the largest shift. Point-of-care
production rises from 5 to 35 publications, while design history and sharing
rises from 11 to 27. This timing helps explain why the groups need to be read
together. Work on sharing and representing knowledge appears earlier in the
corpus, while research on cross-site exchange and hospital production is
concentrated in later periods. Figure \ref{fig:org84cc270} gives the counts
for each period.

\FloatBarrier

The period from 2020 through 2026 contains 147 of the 240 publications. This
number reflects the growth of federated learning and hospital production within
the corpus. It also depends on the search date and the terms used. The map
therefore describes the selected collection, but it cannot measure the size of
a field or support a claim by itself. The findings that follow come from
interpretation of the publications.

\subsection{How Knowledge Moves in Hospital Work}
\label{sec:org9e6fc72}

Knowledge moves through relationships before it moves through software.
Studies of hospital discharge show that staff must translate knowledge across
professions and organisations
\citep{CorpusWaring2014EthnographicStudyKnowledge}. Reviews of professional
networks reach a related conclusion: a connection helps only when roles and
the work carried by the connection are clear
\citep{CorpusCunningham2011HealthProfessionalNetworks}.

Communities of practice provide time and a social setting in which staff can
make experience usable by others. Their operation depends on how members
define a shared problem and who can take part
\citep{CorpusKislov2011CollaborationsLeadershipApplied,CorpusPyrko2016ThinkingTogetherWhat}.
Digital communities can extend this contact, but participation does not by
itself establish the status of the knowledge being shared
\citep{CorpusRolls2016HowHealthCare}.

Hospital knowledge-management studies report recurring organisational
conditions. Staff need access to colleagues and time to explain a case.
Management support affects whether knowledge leaves one team. Trust affects
whether another team acts on it
\citep{CorpusKaramitri2015KnowledgeManagementPractices,CorpusAlmashmoum2023FactorsThatAffect}.
These findings make knowledge sharing a work process rather than a transfer
button.

The same literature gives less attention to the identity of the shared
content. A protocol, an explanation, and an informal lesson can all count as
knowledge sharing. Most studies measure access, participation, or reported
use. They do not require a later result to point back to the version that
informed the work. This leaves the link between a local use and its source
version under-specified. Any lifecycle that closes this gap must retain the
people and review practices on which sharing depends.

\subsection{Why Hospital Data Do Not Become Shared Knowledge by Themselves}
\label{sec:org35bd0e8}

Hospital data require work before another task can interpret them. Records use
different structures and local codes. Their meaning also depends on when and
why staff entered them. Reviews of semantic interoperability show that a
common syntax cannot resolve every difference in meaning
\citep{CorpusMello2022SemanticInteroperabilityHealth,CorpusLehne2019WhyDigitalMedicine}.

Data quality changes when records move from care to analysis. Missing values
may reflect the care process rather than a measurement error. A field can also
change meaning across sites. Research on reuse warns that such patterns can
introduce bias if analysts treat records as observations made for the same
purpose \citep{CorpusVerheij2018PossibleSourcesBias}.

Learning-health-system studies place this problem inside an organisational
cycle. Data can identify a question, but people must decide what the result
means for practice. Reviews find that many programmes build data capability
without closing the path from analysis to later care
\citep{CorpusEnticott2021LearningHealthSystems,PlattEtAl2020LHSReview}.
Leaders also report that governance and staff capacity shape what a learning
system can sustain \citep{CorpusEnticott2020LeadersPerspectivesLearning}.

FAIR data practices and integration pipelines improve access to hospital
records. Their implementation still requires local definitions and data
stewardship \citep{CorpusRosinach2022ApplyingFairPrinciples}. These mechanisms
make data available for a new use. They do not state which reviewed medical
knowledge should change as a result. The review therefore separates source
data from the artifact whose status may change.

\subsection{What Federation Already Provides}
\label{sec:orgf25c412}

Federation is established prior work in health informatics. Distributed
health-data networks send an analysis to sites that retain their source
records \citep{Brown2010Distributed}. Sentinel uses this pattern for
surveillance \citep{PlattEtAl2012MiniSentinel}. SHRINE supports queries across
separate clinical research repositories
\citep{CorpusMcMurry2013ShrineEnablingNationally}.

Common data models reduce the work needed to run one analysis at several
sites. The Observational Medical Outcomes Partnership model supports
transformation into a shared representation
\citep{VossEtAl2015OMOP}. The Observational Health Data Sciences and
Informatics network applies common methods to those transformed records
\citep{HripcsakEtAl2016OHDSI}. Recent reviews show that common data models
continue to differ in scope and implementation
\citep{CorpusFinster2025CommonDataModels}.

Federated learning exchanges model updates rather than source cases
\citep{RiekeEtAl2020Federated}. This mechanism can support joint model
development, but it does not remove differences in populations or local
workflows. Receiving sites still need local validation and authority over use.

Federated work also creates approval tasks. A site must map its records and
review a request before it participates. Studies of distributed research
infrastructures show that local approval can become part of the execution path
\citep{CorpusKombeiz2023LocalApprovalProcesses}. Federation therefore joins
technical exchange with agreements between organisations.

Prior work shows what federation already provides. It can coordinate
work across organisations while each site retains its source records. The
objects exchanged in this literature are queries, results, or model updates.
The reviewed studies do not describe the exchange of a governed medical
knowledge artifact together with feedback linked to the version used at a
site.

\subsection{Medical Knowledge Artifacts as Units of Change}
\label{sec:orga4d053c}

Research on computable biomedical knowledge supplies a unit that software can
store and apply. The Knowledge Object Reference Ontology defines a knowledge
object through its functions and associated roles
\citep{FlynnEtAl2018KORO}. Work on sharable computable knowledge adds
requirements for local validation and continued management
\citep{LehmannDowns2018Desiderata,AdlerMilsteinEtAl2018ManagingCBK}.

Repositories make such objects discoverable. Surveys show support for
metadata, provenance, and contributor identity, but repository functions vary
\citep{PlattEtAl2022Repositories}. Infrastructure guidance adds services for
access and stewardship \citep{McCuskerEtAl2022Infrastructure}. Research on
clinical knowledge interoperability also shows how organisations can exchange
process definitions \citep{CorpusLario2024BusinessProcessManagement}.

Provenance standards explain where a resource came from. PROV links an entity
to the activity and agent involved in its production
\citep{MoreauEtAl2015PROV}. PAV adds relations for authorship, curation, and
versions \citep{Ciccarese2013PAV}. These relations preserve change history. They do
not decide whether a medical change should be approved.

Living guidelines show how planned review can change a recommendation.
Process evaluations report the work needed to monitor questions and publish
updates \citep{CorpusTurner2022AustralianLivingGuidelines}. Yet an update
process need not connect every local use to the exact recommendation version
that informed it. The reviewed processes therefore leave this relation
under-specified.

The synthesis therefore treats improvement as a reviewable claim. A proposed
successor states what changed and why the change should better serve the
artifact's purpose. Reviewers compare that claim with records from use. A
larger version number alone does not establish improvement.

\subsection{Medical Device Knowledge Artifacts and Computer-Aided Design History}
\label{sec:orgeb28f5c}

A \emph{medical device knowledge artifact} is a medical knowledge artifact used to
design or make a medical device. When the content includes computer-aided
design, the artifact preserves the editable design's history. It also links
the selected version to the production record for each use.

Medical three-dimensional modelling begins before a printable file exists.
Staff select source images and segment the structures needed for the task.
Software then converts the result into a surface model. Comparisons of
segmentation workflows show that software choice and user decisions can change
the resulting geometry
\citep{CorpusKamio2020DicomSegmentationStl,FogarasiEtAl2022Segmentation}.

The design process adds another set of choices. Engineers may repair the mesh
or add features needed for manufacture. A design-control account can connect
those decisions to verification
\citep{CorpusHollister2016IntegratingImageBased}.

The search found two ways to share design-related material. Open Knee(s)
publishes models with related digital assets
\citep{CorpusChokhandre2022OpenKneeS}. Open-Full-Jaw publishes a dataset and a
pipeline for model construction
\citep{CorpusGholamalizadeh2022OpenFullJaw}. These resources support reuse and
reproduction. They do not provide a shared review cycle for hospital
adaptations.

Document control addresses another part of the problem. A point-of-care
solution can assign identifiers and approval states to files used in a
printing service \citep{CorpusLohss2023CustomizableDocumentControl}. This
supports local control. The mapped literature gives less attention to branch
relations, change rationale, and reuse of editable designs across hospitals.
This result identifies an under-specified research relation. It does not imply
that no hospital uses file-version tools.

File sharing and artifact improvement are therefore different operations. A
site can copy a surface model without learning why its geometry changed. A
medical device knowledge artifact keeps the design relation and the review
decision with that model. This structure lets another site inspect a
candidate version without treating the file as an approved device design.

\subsection{Point-of-Care Manufacturing as a Test Case}
\label{sec:org0c39d22}

Point-of-care manufacturing makes the need for design history visible because the
hospital also performs production work. Hospital programmes must define who
accepts a request and who prepares the design. They must also assign
responsibility for inspection and release
\citep{CorpusHellman20233dPrintingHospital,CorpusHaro2021PointCareManufacturing}.

Production quality depends on the local process. Equipment and material
choices affect the made item. Staff training affects how the process is
carried out. Reviews of quality assurance show that sites use different
checks and reporting methods
\citep{CorpusSchulze2024QualityAssurance3d,ChristensenRybicki2017Printing}.

Regulatory work places document control and responsibility around this
process. Studies of hospital printing discuss how intended use and production
setting affect the applicable route
\citep{CorpusBeitler2022InterpretationRegulatoryFactors,CorpusPaxton2023NavigatingIntersection3d}.
Surveys also show that hospitals differ in staff capacity and production scope
\citep{VallsEsteveEtAl2024PoC}.

These findings separate two decisions. Artifact review asks whether a proposed
design change should alter shared knowledge. Product release asks whether one
made item meets its intended use. A production observation may inform both
decisions, but the authorities and records differ.

Point-of-care manufacturing therefore shows which knowledge may be shared and
which decisions remain under local control. A site may share design history
with another organisation, but patient records remain at the care site. The
same site also decides whether to release the item it made. The reviewed
studies do not specify how a report from local manufacture should identify the
artifact version that informed it. They also do not show how reviewers across
sites could distinguish a reusable change from a case-specific adaptation.

\subsection{Integrated Findings}
\label{sec:org72692fc}

Taken together, the six literature groups reveal a missing connection rather
than a need for another health-data network. Each group explains one part of
the path from using medical knowledge to changing it for later use. The
publications do not join those parts into one process. Such a process would
begin when staff use a known version and end when the people responsible for it
decide whether the shared knowledge should change. Table
\ref{tab:org1ab2c75} shows what each group explains and what still needs
to be connected.

\begin{longtable}{@{}p{0.25\linewidth}p{0.25\linewidth}p{0.40\linewidth}@{}}
\caption{\label{tab:org1ab2c75}What each literature group explains and what remains to be connected.}
\\[0pt]
Literature group & What prior work explains & What remains to be connected\\[0pt]
\hline
\endfirsthead
\multicolumn{3}{l}{Continued from previous page} \\[0pt]
\hline

Literature group & What prior work explains & What remains to be connected \\[0pt]

\hline
\endhead
\hline\multicolumn{3}{r}{Continued on next page} \\
\endfoot
\endlastfoot
\hline
Hospital knowledge sharing & How staff share what they know through work and professional networks & Identify the reusable knowledge and record which version staff used\\[0pt]
Hospital data and learning systems & How organisations prepare care records for analysis and use the findings & Keep case records separate from the medical knowledge artifact under review\\[0pt]
Federated health infrastructures & How sites exchange queries, results, or model updates while records stay local & Identify the artifact version and its review status across sites. Link each site's report about its use to that version.\\[0pt]
Computable medical knowledge & How repositories store rules or guidelines and record their change history & Link a record of use to a decision about whether the artifact should change\\[0pt]
Medical computer-aided design & How staff create design files from medical images and record design history & Preserve why a design changed and which changes passed review\\[0pt]
Point-of-care production & How hospitals assign responsibility for making and releasing a device & Keep review of shared design knowledge separate from release of the made device\\[0pt]
\end{longtable}

Closing this gap requires two links. One connects each use to the artifact
version that informed it. The other connects a proposed change to a review
decision. An application record preserves the first link by naming the version
used. It also records what staff did and what result they observed. If that
result suggests a change, a proposal names the version that staff used and
explains what should change. The people responsible for the artifact then
decide whether the proposal becomes a version for future use. A local result
can therefore inform review without changing the shared artifact by itself.

This process can operate within one hospital. Staff can link each use to a
version. People responsible for the artifact can review a proposed change and
issue a new version. An exchange infrastructure extends the process across
sites by helping them find artifacts and send proposals to the people
responsible for them. Each site continues to control its records and its local
use.

The findings also set an order for implementation and evaluation. A first study
can test whether one use remains linked to the artifact version that informed
it. A study across sites can then measure the work required for review. It can
also check whether the record shows who made each decision and why. A
point-of-care manufacturing study can test whether the process links a made
device to its design version while keeping design review separate from product
release. Studies of clinical outcomes and regulatory conformity can follow as
their own lines of evaluation. These findings support the requirements for the
exchange infrastructure proposed in the next section.

\section{Conceptual Model and Derived Requirements}
\label{sec:orgb042533}

\subsection{Gap Identified by the Review}
\label{sec:org32158e6}

The literature review identifies a missing step rather than a need for
another repository. Repositories store medical knowledge. Provenance models
record how a version arose
\citep{AlperEtAl2021Metadata,Ciccarese2013PAV,PlattEtAl2022Repositories}.
Federated networks let sites work together while source records remain under
local control \citep{Brown2010Distributed,RiekeEtAl2020Federated}. These
mechanisms do not provide a common path from the use of one artifact version
to a reviewed change in that artifact.

The missing path begins when a site applies an artifact. Staff need to record
which version informed the work and what followed. If the result suggests a
change that may matter outside the case, authorised staff must state that
claim without sharing the patient record. The proposal must then reach the
people who govern the artifact. Their decision must return to sites that may
use it.

Point-of-care manufacturing makes the missing path visible. A hospital may
adapt a design and make an item from it. The hospital decides whether to
release that item. A separate review decides whether the design change should
alter the shared medical device knowledge artifact. One decision cannot stand
in for the other.

\subsection{Proposed Exchange Infrastructure}
\label{sec:org6820503}

We therefore propose an exchange infrastructure for the continuous
improvement of medical knowledge artifacts. It connects each application to
the artifact version that informed it. It then carries an authorised report
or change proposal to the people who govern that artifact. Their decision
returns as a new version or a change in status.

The model separates three linked records. A case record supports care and
remains at the care site. A medical knowledge artifact contains content that
people may assess and reuse. It states its purpose and scope. It also records
its change history and review status. An application record names the version
or local branch used and records what followed.

These records support a review process without turning each reported result
into accepted knowledge. If an application suggests a change, authorised
staff prepare a proposal. The proposal names its parent and explains the
difference. Reviewers may accept a successor or retain a version for one
setting. They may reject the proposal or change the status of an earlier
version. The proposal cannot change the accepted artifact by itself.

Federation extends this process across organisations, but it does not create
the improvement decision. A receiving site decides whether to use an artifact
and what information it may return. The source steward decides whether a
returned proposal changes the shared artifact. The exchange infrastructure
therefore carries artifacts and review records between sites. It does not
transfer authority between them or require one database.

\subsection{Derived Requirements}
\label{sec:orgcd6e3b7}

\begin{longtable}{@{}p{0.34\linewidth}p{0.56\linewidth}@{}}
\caption{\label{tab:org01db649}Propositions and requirements derived from the literature review.}
\\[0pt]
Proposition & Derived requirement\\[0pt]
\hline
\endfirsthead
\multicolumn{2}{l}{Continued from previous page} \\[0pt]
\hline

Proposition & Derived requirement \\[0pt]

\hline
\endhead
\hline\multicolumn{2}{r}{Continued on next page} \\
\endfoot
\endlastfoot
\hline
Context makes an artifact assessable & Record its purpose and scope. Preserve its origin, change history, and review status.\\[0pt]
Feedback needs a version-and-application link & Name the version or branch used. Record the conditions of use and what followed.\\[0pt]
A variant can be an improvement outcome & Keep an adaptation linked to its parent and state where it applies. Do not overwrite the source.\\[0pt]
Review creates governed status & A named person decides whether a proposal becomes a correction, variant, or successor. Record the reason and affected versions.\\[0pt]
Local governance precedes federation & Let each organisation select, apply, branch, and review an artifact without depending on a network connection.\\[0pt]
Federation adds reach and work & Preserve identity across sites. Carry authorised feedback and status notices, while measuring conflicts and governance workload.\\[0pt]
\end{longtable}

The propositions in Table \ref{tab:org01db649} turn the findings of
the review into requirements that can later be tested. They do not assume
that every change is an improvement. They require a link between use and
version because a result cannot inform review when its source is unknown.
They also preserve local variants because improvement may depend on the
setting.

The local process begins when staff select an artifact after checking its
purpose and status. A local adaptation creates a branch rather than rewriting
the source. The application record keeps the link to the version used. It
also retains an unfavourable or inconclusive result. A review decision closes
the process and notifies known recipients. A new file alone does not
establish improvement.

The lifecycle must work within one organisation before it can work across
sites. Loss of a connection must not stop local use or review. Federation
widens discovery and lets more sites return reports. It can also produce
competing branches and more work for reviewers. When sites propose different
changes, reviewers keep their scopes and decisions distinct. Corrections and
withdrawals must reach known recipients.

A shared vocabulary lets sites interpret the records in the same way. It
states purpose and scope. It also represents version relations and review
status. Local systems may retain more detail. Automated services may help
people find artifacts and check links. They may also prepare a draft. Each
output records its source and the person's response. A person decides whether
an artifact applies. The named authority decides whether a proposal changes
accepted knowledge or artifact status.

Point-of-care manufacturing adds a second governance decision. A medical
device knowledge artifact links design to production. The editable
computer-aided design source remains linked to each derived manufacturing
file and the produced item. Patient records and inspection records remain at
the care site. Staff there decide whether to release the item
\citep{CorpusLohss2023CustomizableDocumentControl,CorpusHellman20233dPrintingHospital}.
Artifact review cannot release a device or establish conformity. The
responsible organisation still selects the regulatory route and completes the
required controls
\citep{EuropeanUnion2017MDR,MDCG2021CustomMade,MDCG2023InHouse}.

\subsection{Role of Open-Source Hardware in Medical Device Knowledge Artifacts}
\label{sec:org4323fed}

A shared design history does not give another site the right to change a
design. If sites are expected to inspect and adapt a computer-aided design
source, that source must be released as open-source hardware. The licence must
also let them return the changed design. The release must include the form used
to edit the design. It must permit modification and redistribution
\citep{NiezenEtAl2016OpenMedicalHardware,OSHWAOpenHardwareDefinition}.
Open-source hardware is therefore a precondition for a cross-site improvement
process with the same reuse rights at each site. A file that can be downloaded
or viewed does not meet this condition by itself.

Research on open hardware explains why access to the source matters. Pearce
shows that researchers can reproduce equipment and adapt it when the design
files are open \citep{Pearce2012OpenHardware}. Laverty and colleagues report
that institutions copied an open measurement system and improved its design
\citep{LavertyEtAl2013OpenPMU}. In both cases, the source files let another
group act on the design rather than rely on a finished device. A design history
can then record which change came from which version.

Medical use adds controls that an open licence cannot provide. Niezen and
colleagues argue that open-source hardware can support faster work on medical
devices \citep{NiezenEtAl2016OpenMedicalHardware}. Chagas and colleagues show
how shared designs allowed people in many places to contribute during a health
emergency \citep{ChagasEtAl2020OpenHardware}. Chagas and colleagues also
identify the need for testing and documentation. The licence enables design
work. It does not approve a device or establish that a version is safe.

The exchange infrastructure should record the hardware licence for every design
version. A reciprocal licence from the CERN Open Hardware Licence family fits
the proposed improvement process
\citep{AyassSerrano2012CERNOHL,CERN2020OHLv2,OSHWASharingBestPractices}.  Its
weak and strong reciprocal forms require covered design changes to remain
available under the same licence when their terms apply. Under the weak form,
this duty stays with the modified part when it becomes part of a larger
design. The strong form also applies the duty to the larger design. A steward
can choose the form that fits the design. The exchange record must retain that
choice and flag a proposed merge when the licences conflict. A Creative Commons
licence may cover text or images, but it should not serve as the sole licence
for the hardware design source because Creative Commons does not recommend its
licences for hardware \citep{CreativeCommonsFAQ}.

Exchange could still work when a design cannot be released as open-source
hardware. Sites may share its identity and review status. They may also send
change notices. Access to the source can depend on a contract or permission from
the owner. This approach makes each reuse decision depend on the rights held by
the receiving site. Some branches cannot return to the shared design, and a
steward cannot merge them without checking those rights. The exchange
infrastructure can represent such limits, but it cannot provide one improvement
process for all sites.

\section{Illustrative Point-of-Care Manufacturing Scenario}
\label{sec:org782cc0a}

A point-of-care FDM workflow makes the effect of each knowledge decision visible. The
clinical need leads to a design, and the design leads to a physical item. An error in
the design history can therefore affect both later knowledge and the item produced in the current
case. In the setting that motivates this paper, no single clinician-facing workflow
connects all of these decisions. Case information remains in clinical systems, while
CAD work and printer records arise elsewhere. Quality documentation must then be
assembled from those separate systems. The following scenario explains how the model
could address that problem; it is not an implementation or clinical result.

\begin{figure}[H]
\centering
\includegraphics[width=0.98\linewidth]{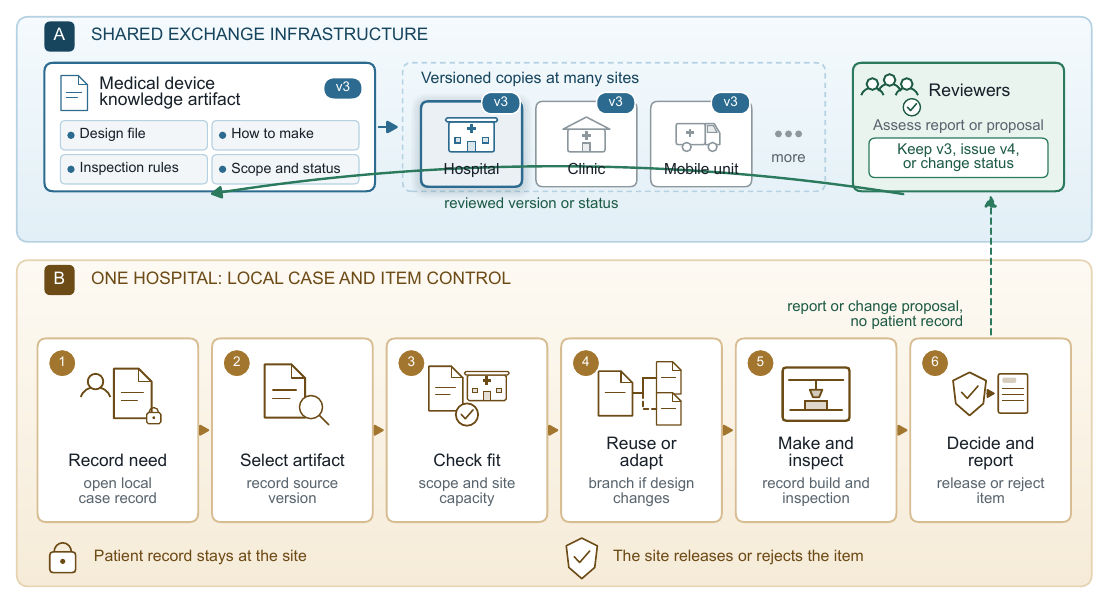}
\caption{\label{fig:org33bae34}Exchange and local use of a medical device knowledge artifact. Panel A shows version 3 distributed to a hospital, clinic, and mobile unit. Panel B follows one hospital from recording a need to deciding whether to release the item. The site keeps patient records and item-release authority. It may send a report or change proposal for review. Reviewers may retain the version or issue a successor. They may also change its status.}
\end{figure}

Figure \ref{fig:org33bae34} separates shared exchange from local care and
production. Panel A shows version 3 of a medical device knowledge artifact. Its
record contains the design file and explains how to make the item. It also
states how to inspect the result. The scope and status guide selection at each
site. The exchange infrastructure supplies versioned copies to a hospital, a
clinic, and a mobile unit.

Panel B follows one hospital from a local need to the item decision. Staff open
a local case record and select an artifact. They record the source version
before assessing whether the artifact fits the case and whether the site can
make it. A design change creates a branch. Staff then make and inspect the item.
The site decides whether to release or reject it. Throughout this process, the
patient record remains at the site. The site may send a report or change
proposal without sending that record. Reviewers assess the submission. They may
keep version 3 or issue version 4. They may instead change the artifact's
status. The reviewed version or status then returns through the exchange
infrastructure.

The scenario begins with a governed \emph{medical device knowledge artifact} for a
patient-specific FDM device. The artifact states its intended purpose and the limits
within which the source organisation has accepted it. Its reusable content includes an
editable CAD model and the rationale for the design. A manufacturing extension
identifies the process specification that supports that design. Inspection criteria
explain how staff judge the produced item. Evidence references and review status tell a
receiving team what has been established. Patient images and treatment notes remain in
the local case rather than in the shared artifact.

A clinician first records the new need and searches for an artifact with a matching
purpose. The clinician and an engineer assess whether the artifact's scope covers the
case. They also compare its manufacturing assumptions with the resources available at
their site. The application record identifies the selected version and preserves the
reason for the decision. If the geometry must change, the engineer creates a branch
from that version. The branch is not presented as an approved improvement.

Within the medical device knowledge artifact, the design history connects the design decision
to fabrication. The branch records the changed dimensions in the editable source model
and identifies who made the change. A derived mesh points back to that source version.
The build preparation then links its machine instructions to the mesh and to the process
specification used to generate them. These relations let a reviewer distinguish a
design change from a change introduced during file conversion. They also prevent a
later user from treating the last exported file as the complete design history.

Fabrication creates an application record rather than an automatic artifact update.
The production system records which machine executed the build and which material lot
it used. Process observations show whether the build remained within the specified
window. Inspection attaches measurements to the produced item and records any
nonconformity. Qualified staff decide whether to release or reject that item. Their
decision remains part of the local quality process, even when the same evidence later
informs artifact review.

The application outcome may reveal knowledge that matters beyond the case. An observed
fit problem could show that the design's stated applicability is too broad. A repeated
result could instead support the existing process specification. Authorised staff
prepare an evidence submission that states the claimed implication. They remove
patient information that is not needed for review and retain a controlled link to the
source. The submission can propose a new variant or ask reviewers to change the
artifact's scope. It does not change the accepted version by itself.

Review turns the submission into a governed knowledge decision. Reviewers may conclude
that the local branch applies only to the originating case. They may accept it as a
variant for a defined setting, or they may promote a successor after further evidence.
A negative outcome can lead to a correction or withdrawal. The record also allows
reviewers to leave the current artifact unchanged when the outcome does not support a
general claim. In every case, review of shared knowledge remains separate from release
of the physical item.

Federation allows another organisation to enter this lifecycle without surrendering
local responsibility. A site discovers the approved medical device knowledge artifact
and inspects its evidence before importing it. Local users create their own application
record and retain the patient case under local custody. After use, the site may return
an approved evidence submission or a revision candidate. The federation preserves the
source version so that the governing reviewers can interpret the feedback. If they
later correct or withdraw that version, the exchange infrastructure sends a status
notice to the receiving site.

An optional regulatory profile can derive controlled documents from these records. For
an in-house pathway, it could help organise the intended purpose and the manufacturing
description required under Article 5(5). It could also connect the device identifier to
experience from use \citep{EuropeanUnion2017MDR,MDCG2023InHouse}. A different device
route would require a different profile. The generated package would remain a draft
inside the organisation's quality management system. It could improve traceability, but
it could not decide classification or demonstrate compliance on its own.

\section{Discussion}
\label{sec:orge4adde7}

Healthcare staff reuse medical knowledge when they follow a protocol or make a
device from an existing design. Improving that knowledge requires more than
making it available for reuse. An organisation must connect the version used
and the result that followed to a decision about whether the knowledge should
change. The publications reviewed in this paper explain parts of this process,
including local use and exchange between sites, but do not describe how those
parts connect. The model supplies the connection through two records. An
application record identifies the version used and states what happened during
the work. A review decision records whether the people responsible for the
protocol or design accepted a proposed change. Together, these records let
later users trace an accepted change back to the work that started it.

Learning health systems seek to improve care by learning from earlier work.
Research in this field describes a cycle that begins with information from
practice. People analyse that information and use the findings to guide later
work \citep{InstituteMedicine2007Learning,FriedmanEtAl2016LHS}. The cycle does
not require each use of a protocol or design to name the version that informed
it. Without that link, reviewers cannot tell whether a reported result concerns
the current version or an earlier one. The proposed model adds the link through
an application record. Staff may use the record to propose a change, but the
people responsible for the protocol or design decide whether to accept it.
This allows use to inform change without making every report a new version.

Repositories and provenance systems answer a different question: what is the
knowledge item, and how did it change? A repository helps people find a
protocol or design. A provenance record identifies its source and connects one
version to another
\citep{AlperEtAl2021Metadata,Ciccarese2013PAV,PlattEtAl2022Repositories}.
This history can show that one file came from an earlier file. It does not show
whether the people responsible for the knowledge reviewed and accepted the
change. The proposed model connects the change history to that decision. It
therefore distinguishes an edited file from knowledge approved for later use.

Federation addresses how organisations work together without placing their
source records in one database. Health networks use this arrangement to
coordinate analyses across sites while each site keeps its records
\citep{Brown2010Distributed,HripcsakEtAl2016OHDSI}. This paper does not claim
federation as a new method. It proposes what sites should exchange when they
want to improve shared knowledge. The exchange infrastructure would carry a
version of a protocol or design and state whether reviewers had accepted it. A
site could return a report that names the version it used. Local use and review
could continue when the site cannot reach the network. Adding sites may produce
more proposals and more conflicts between versions, so an evaluation must
measure review work as well as network reach.

Improvement does not require every site to adopt the same version. Reviewers
may accept a proposed change as the next version for later use. They may keep
it as a variant that applies to one setting. They may instead reject it and
retain the current version. Each outcome can be appropriate when the decision
states its reason and where the version applies. Counting updates would miss
this distinction. An evaluation should ask whether users select a version that
fits their setting and whether they can trace its status to the uses considered
during review. Review time and missing record links show the work required to
maintain this process.

Using medical knowledge and changing shared knowledge are separate decisions.
Staff at a care site decide whether a protocol or design fits a case. The people
responsible for that knowledge decide whether a proposal changes the version
available to later users. If the site makes a device, qualified staff make a
third decision about releasing the made item. One decision cannot stand in for
another. An interface must show who made each decision and what that decision
allows. Software may flag a missing link or records that conflict, but a named
person remains responsible for the response. A test must therefore examine what
users do when software gives a wrong suggestion, not only how often the
software is correct.

Sharing a medical device design also requires permission to change its editable
file. Downloading the file gives access, but it does not always grant the right
to edit it or return a changed version. An open-source hardware licence can
grant those rights
\citep{NiezenEtAl2016OpenMedicalHardware,OSHWAOpenHardwareDefinition,CERN2020OHLv2}.
The licence does not approve the design or establish that a made item is safe.
When access depends on a contract, sites may still exchange the design's
identity and state whether reviewers accepted it. A site may lack the right to
submit its changed version for shared review. An implementation must therefore
keep the licence with each version and warn reviewers when licence terms
conflict.

Point-of-care manufacturing tests the model in a process that creates design
and production records. In this setting, a hospital or clinic makes a device
near the place of care. The design has a version, and the made item has
production records. These records make it possible to trace a local use back to
the design and to keep design review separate from release of the item
\citep{CorpusLohss2023CustomizableDocumentControl,CorpusHellman20233dPrintingHospital}.
Other forms of medical knowledge may not have this structure. A guideline gains
acceptance after experts review the research, while a predictive model depends
on the setting in which people test it. Some knowledge remains part of a
person's practice and cannot be managed as a file. Research must therefore
compare a device design with another form of medical knowledge before applying
the model beyond the device case.

The concept is designed to apply across health organisations because it defines
a review process rather than one software product or production workflow. Under
the model, a hospital could share an accepted version of a medical device
knowledge artifact while the patient record remains at the care site. A
receiving organisation could record how it used that version and return a
report or proposed change. The people responsible for the artifact would then
decide whether the shared knowledge should change. The same process could
support a clinical protocol or another form of recorded medical knowledge whose
use can be reviewed. The concept therefore reframes exchange between hospitals:
instead of sending a file without its change history, organisations preserve
how each version was used and why reviewers accepted or rejected a proposed
change. Implementation studies can examine how this process fits current work
and connects to existing software. Each implementation can define its security
requirements based on where it runs and the threats it faces. Later studies
can determine whether the process improves care.

Keeping patient records at the care site limits what must pass through the
exchange infrastructure, but it does not make every shared report anonymous.
For example, the shape of a device made for one patient may identify that person
after names and record numbers are removed. Each site must determine whether it
has a legal basis to share such a report. Federation cannot make that decision
or assign responsibility between organisations. The model's regulatory profile
is a way to arrange records for a regulatory route. It cannot show that a device
conforms to the law or give another organisation permission to make it. Those
decisions depend on the device and the regulatory route
\citep{EuropeanUnion2017MDR,MDCG2021CustomMade,MDCG2023InHouse}. Any study
across sites must assign these responsibilities before case-derived reports are
exchanged. These limits do not remove the learning value of records linked to a
medical device knowledge artifact. When a case-derived report cannot leave the
care site, staff can still link it to the artifact version used and examine why
the use succeeded or failed. Clinicians and engineers can use the record to
teach how staff selected the design. If staff changed it, the record can explain
why. It can also show how staff reached review and release decisions. When a
site determines that a report can be shared after patient-specific information
is removed or generalised, the report can support learning in other
organisations. The model therefore supports case-based learning within a site
and shared learning across sites when disclosure is permitted.

A first study should follow one patient-specific device case from the clinical
need to a decision about shared knowledge. The clinician would record the need
and review earlier cases held at the hospital to see how staff addressed it.
Those cases could point to a medical device knowledge artifact used before. The
clinician and an engineer would then inspect the current artifact version and
decide whether its purpose covers the new case. If the geometry must change,
the engineer would create a branch and record why. The engineer would derive
the manufacturing file from that version or branch. Production staff would use
a named process specification and record the inspection result. Qualified
staff would then decide whether to release the made item. The application
record would link the local case to the artifact version. It would also link
that version to the production and release records. After use, staff would
record what happened and decide whether the result concerns only the case or
supports a proposal to change shared knowledge. Reviewers should be able to
reconstruct this path without receiving the patient record. The study should
also test whether the people named in each record hold the authority that the
model assigns to them. Once this process works within one hospital, a second
site could apply the same parent version to another case and propose a different
change. Reviewers would then decide whether either proposal applies beyond its
case and whether the sites need separate variants. A later correction or
withdrawal would test whether notices reach both sites. The study would measure
whether each made item remains linked to its design version and production
process. Later studies would assess device quality and clinical outcomes.

\section{Conclusion}
\label{sec:orgaa6d656}

This paper proposes an exchange infrastructure that governs changes to medical
knowledge artifacts/, which are named and versioned units of reusable and each
proposed change to a review decision. A \emph{medical device knowledge artifact}
applies the same model to device design and production. The contribution defines
what organisations exchange through federation and how shared knowledge changes
after use.

The integrative review screened 910 unique records and mapped 240 publications
across six groups. It then examined 72 publications through close reading. The
comparison exposed a gap between learning at one site and changing knowledge for
later users. Prior work gives organisations ways to store versions and exchange
information. It does not define how a result from one version becomes a proposal
with a stated scope or how the review outcome returns to sites that may rely on
it. The proposed model turns this gap into a sequence of records and
decisions. An application record states which artifact version was used and what
happened during that use. A change proposal names the artifact version on which
it is based and describes what staff want to change. An authorised review
assigns the proposal a status and scope. These relations determine what the
exchange infrastructure carries between sites. They show whether a change
remains at one site or has been approved for use at other sites. They also link
the review decision to the use that led staff to propose the change.

The insight is that federation can distribute production knowledge and
its review cycle. A medical centre with digital fabrication capabilities could
serve as a user of an accepted artifact and a source of proposed changes.
Discovery becomes a capability-matching task. The artifact's medical purpose
must fit the case. Its production requirements must fit local equipment and
inspection methods. In disaster response, a mobile medical centre could
receive governed production knowledge and fabricate near the point of need.
The patient record remains at the care site. The centre can return a
version-linked report after production. The exchange creates traceability from
design to item and from use to review. Artifact reviewers govern shared status,
while the care site governs item release.

Further studies could implement the model in a medical centre with digital
fabrication capabilities. A first study could trace one device from artifact
discovery to product release and later artifact review. A federated study could
connect the centre to a mobile disaster-response fabrication unit. The mobile
unit could receive an accepted version before deployment and continue local
work during a network interruption. After reconnection, it could return its
application record and receive a correction or withdrawal notice. This
sequence would test local continuity and status synchronisation. As more sites
contribute, review capacity becomes a scaling concern. Studies could measure
review time and resolution of competing branches. Later evaluations could
examine clinical outcomes and support for regulatory work. Together, these
studies could show how federated medical centres turn distributed fabrication
experience into shared medical knowledge while patient records and release
authority remain local.

\section{Acknowledgement}
\label{sec:orgbee5da3}

The project ``Fab City---Decentralized Digital Production for Urban Value
Creation'' is funded by dtec.bw---Digitalization and Technology Research Center
of the Bundeswehr. dtec.bw is funded by the European Union---NextGenerationEU.

\section{Conflicts of Interest}
\label{sec:orgf52a25d}

The authors declare no conflicts of interest.

\section{Generative AI Use}
\label{sec:org8439f21}

During preparation of this manuscript, the authors used GPT5.6 Sol to assist
with language refinement, clarity, and readability. All AI-assisted output was
critically reviewed, edited, and verified by the authors.

\pagebreak
\parindent=0pt

\bibliographystyle{unsrtnat}
\nocite{CorpusBennett2024ImprovingPatientsExperiences,CorpusAlmashmoum2023FactorsThatAffect,CorpusColladon2022BoostingAdviceKnowledge,CorpusShaw2022ImplementationVirtualCommunities,CorpusAbbate2022InvestigatingHealthcare4,CorpusKosklin2022KnowledgeManagementEffects,CorpusColnar2022RoleInformationCommunication,CorpusHenshall2022UnderstandingClinicalDecision,CorpusMayr2021FocusDietQuality,CorpusKaramitri2020DevelopmentValidationKnowledge,CorpusWu2020DoWorkEngagement,CorpusYan2020HowHospitalsMainland,CorpusShateri2020InvestigatingMediatingRole,CorpusAgrifoglio2020UnderstandingKnowledgeSharing,CorpusKaramat2019DevelopingSustainableHealthcare,CorpusKaramat2019PromotingHealthcareSustainability,CorpusConstance2019SupervisoryJusticeOrganizational,CorpusKaramat2018BarriersKnowledgeManagement,CorpusSamad2018TheoryPlannedBehavior,CorpusShahmoradi2017KnowledgeManagementImplementation,CorpusRolls2016HowHealthCare,CorpusPyrko2016ThinkingTogetherWhat,CorpusKaramitri2015KnowledgeManagementPractices,CorpusWaring2014EthnographicStudyKnowledge,CorpusGebretsadik2014KnowledgeSharingPractice,CorpusTabrizi2014ModelsDescribingKnowledge,CorpusRowley2014ProtocolExplorationKnowledge,CorpusShibuya2013ApproachMedicalKnowledge,CorpusKislov2013BoundaryDiscontinuityConstellation,CorpusStewart2012ApplyingSocialNetwork,CorpusZach2012CapturingInformationNeeds,CorpusWu2012ExaminingKnowledgeManagement,CorpusBarnett2012GeneralPracticeTraining,CorpusKislov2011CollaborationsLeadershipApplied,CorpusCunningham2011HealthProfessionalNetworks,CorpusKothari2011LessonsBusinessSector,CorpusJacobs2011MediatingEffectKnowledge,CorpusHamornik2010KnowledgeSharingMedical,CorpusJacobs2008OrganisationalCultureHospitals,CorpusHsia2006FrameworkDesigningNursing,CorpusFinster2025CommonDataModels,CorpusHaber2025CoreConceptsPharmacoepidemiology,CorpusKim2024DataResourceProfile,CorpusPark2024DevelopmentMedicalImaging,CorpusAbbas2024FederatedLearningSmart,CorpusLazaros2024FederatedLearningNavigating,CorpusPark2023ExploringPotentialOmop,CorpusLee2023FeasibilityStudyFederated,CorpusRaventos2023IncidenceprevalenceRPackage,CorpusKombeiz2023LocalApprovalProcesses,CorpusLuo2022DlmmLosslessOne,CorpusYou2022EstablishmentInternationalEvidence,CorpusPati2022FederatedLearningEnables,CorpusAouedi2022HandlingPrivacySensitive,CorpusYoo2022TransformingThyroidCancer,CorpusDayan2021FederatedLearningPredicting,CorpusMaier2021PatientCohortIdentification,CorpusBiedermann2021StandardizingRegistryData,CorpusParis2021TransformationEvaluationMimic,CorpusLamer2021TransformingAnesthesiaData,CorpusKent2020CommonProblemsCommon,CorpusKaissis2020SecurePrivacyPreserving,RiekeEtAl2020Federated,CorpusRyu2020TransformationPathologyReports,CorpusLamer2020TransformingFrenchElectronic,CorpusShin2019GenomicCommonData,CorpusLynch2019IncrementallyTransformingElectronic,CorpusGruendner2019KetosClinicalDecision,CorpusLai2018ApplyingCommonData,CorpusHong2018PreliminaryExplorationSurvival,HripcsakEtAl2016OHDSI,CorpusYoon2016ConversionDataQuality,CorpusResnic2015CreatingCommonData,VossEtAl2015OMOP,CorpusHernandez2015AdaptableTrialPcornet,CorpusMcMurry2013ShrineEnablingNationally,CorpusSchilling2013ScalableArchitectureFederated,PlattEtAl2012MiniSentinel,CorpusOverhage2011ValidationCommonData,Brown2010Distributed,CorpusYazdinejad2026BreakingInterprovincialData,CorpusYadegari2025UnifiedIotArchitectural,CorpusSaberi2025DataSilosHealth,CorpusAdeshina2025InteroperableItArchitectures,CorpusZiegler2024BridgingDataSilos,CorpusTripathi2024BuildingFlexibleScalable,CorpusWiley2024BuildingVerticallyIntegrated,CorpusWilliams2023StandardizedClinicalData,CorpusPetch2023DevelopingDataAnalytics,CorpusParciak2023FairnessThroughAutomation,CorpusRosinach2022ApplyingFairPrinciples,CorpusMcDonald2022IdentifyingRequisiteLearning,CorpusChen2022NudgingWithinLearning,CorpusSzarfman2022RecommendationsAchievingInteroperable,CorpusMello2022SemanticInteroperabilityHealth,CorpusEllis2022ScienceLearningHealth,CorpusAllen2021RoadmapOperationalizeEvaluate,CorpusEasterling2021ClarifyingConceptLearning,CorpusAsiimwe2021BiobankDataSilos,CorpusEnticott2021LearningHealthSystems,CorpusCumyn2021MetaConsentSecondary,CorpusMcGraw2021PrivacyProtectionsEncourage,CorpusParsons2021SevenPracticesPursuing,CorpusSchleyer2021IndianaLearningHealth,CorpusInau2020InitiativesConceptsImplementation,CorpusEnticott2020LeadersPerspectivesLearning,CorpusHarrison2020MultilevelAnalysisLearning,CorpusAtsma2020UnderstandingUnwarrantedVariation,PlattEtAl2020LHSReview,CorpusSafaeinili2019CfirSimplifiedPragmatic,CorpusLehne2019WhyDigitalMedicine,CorpusJeffries2018DevelopingLearningHealth,CorpusScott2018LearningHealthSystems,CorpusVerheij2018PossibleSourcesBias,CorpusMcLachlan2018HeimdallFrameworkSupporting,CorpusLessard2017ArchitecturalFrameworksDefining,CorpusFoley2017WhatRoleLearning,CorpusEvans2016ElectronicHealthRecords,FriedmanEtAl2016LHS,CorpusMarsolo2015DigitalArchitectureNetwork,CorpusTousidonis2026AcademicPointCare,CorpusLeon2026AdvancedDigitalWorkflow,CorpusSharma2026ClinicalImplementationEu,CorpusTousidonis2026GuidedImplantSurgery,CorpusWendo2025DimensionalAccuracyAssessment,CorpusMarquardt2025MandibularReconstructionOutcomes,CorpusMonalisa2025TransformingDentalCare,CorpusKantaros2024AdditiveManufacturingSurgical,CorpusMaintz2024PatientSpecificImplants,VallsEsteveEtAl2024PoC,CorpusSchulze2024QualityAssurance3d,CorpusJaksa2024StateArtMedical,CorpusDean2024VirtualSurgicalPlanning,CorpusSlavin20233dPrintingApplications,CorpusHellman20233dPrintingHospital,CorpusValls2023AdvancedStrategiesFabrication,CorpusMendonca2023Overview3dAnatomical,CorpusTrung2023EarlyOutcomesTotal,CorpusAlzoubi2023EmpoweringPrecisionMedicine,CorpusPatel2023Medical3dPrinting,CorpusPaxton2023NavigatingIntersection3d,CorpusArmendariz2023WorkflowRoboticPoint,CorpusAuriemma2022AdditiveManufacturingStrategies,CorpusBonacorsi2022CommunicationDecisionSupport,CorpusSalazar2022Comparison3dPrinted,CorpusBicudo2022InstitutionalInfrastructureChallenges,CorpusBeitler2022InterpretationRegulatoryFactors,CorpusSzary2022Role3dPrinting,CorpusHaro2021ConceptualEvolution3d,CorpusDaoud2021EstablishingPointCare,CorpusSchaffarczyk2021HumanCentricRegulatory,CorpusHonigmann2021Inhospital3dPrinted,CorpusHaro2021PointCareManufacturing,CorpusSharma2021QuantitativeAssessmentPoint,CorpusBicudo2021UkSEmerging,CorpusMathew2019FusedDepositionModeling,CorpusOdeh2019MethodsVerification3d,ChristensenRybicki2017Printing,DiPrimaEtAl2016FDA,CorpusChan20153dRapidPrototyping,CorpusBeyer2026SlicerIndependentFramework,CorpusElrefaei2025Customized3dPrinted,CorpusLuo2025SystematicIndividualizedPreparation,CorpusChrz2024CostAffordableMethodology,CorpusRavi2024ImageProcessingIncluding,CorpusDrakopoulos2024ImageMeshConversion,CorpusChoi2024OptimizingDicomFile,CorpusChegini2023TrainingToolClinicians,CorpusLohss2023CustomizableDocumentControl,CorpusBieleda20223dPrintingIndividual,CorpusChegini202263TrainingTool,FogarasiEtAl2022Segmentation,CorpusGhosh2022Clinical3dModeling,CorpusKamio2022FusedDepositionModeling,CorpusFlorkow2022MagneticResonanceImaging,CorpusChokhandre2022OpenKneeS,CorpusGholamalizadeh2022OpenFullJaw,CorpusLi2022PatientSpecificDaily,CorpusRobb2022CurrentPossibleFuture,CorpusTalanki2021ThreedimensionalPrintedAnatomic,CorpusBoedecker2021Virtual3dModels,CorpusChamo2020AccuracyAssessmentMolded,CorpusMamdouh2020Converting2dMedical,CorpusHolmes2020CreationAnthropomorphicCt,CorpusKamio2020DicomSegmentationStl,CorpusCogswell2020IntracranialVasculature3d,CorpusNarita2020UtilizationDesktop3d,CorpusNagassa2019Advanced3dPrinted,CorpusZaid2019CreatingCustomizedOral,CorpusWake2019PatientSpecific3d,CorpusWallner2018ClinicalEvaluationSemi,CorpusPozzi2018SmilingScanTechnique,CorpusPunyaratabandhu20183dModelsOrthopedic,CorpusMohammed2017AdvancedAuricularProsthesis,CorpusHollister2016IntegratingImageBased,CorpusBiglino20153dManufacturedPatient,CorpusNaftulin2015StreamlinedInexpensive3d,CorpusKrauel2015Use3dPrototypes,CorpusKistler2013VirtualSkeletonDatabase,CorpusAntiga2008ImageBasedModeling,CorpusDomingues2026DesignImplementationDmgcdss,CorpusLario2024BusinessProcessManagement,CorpusJaiyesimi2024TherapyStageIv,CorpusJaiyesimi2024TherapyStageIv2,CorpusLabkoff2024TowardResponsibleFuture,CorpusWyatt2023WhichComputableBiomedical,PlattEtAl2022Repositories,McCuskerEtAl2022Infrastructure,CorpusEnglish2022LivingClinicalGuidelines,CorpusTurner2022AustralianLivingGuidelines,CorpusDullabh2022TechnicalLandscapePatient,AlperEtAl2021Metadata,CorpusGavin2021PosterAbstractsFourth,CorpusOliveira2021QualityDocumentationNursing,CorpusWilliams2021SummaryFourthAnnual,CorpusSutton2020OverviewClinicalDecision,CorpusMahadevaiah2020ArtificialIntelligencebasedClinical,CorpusAlper2020DevelopingMetadataCategories,CorpusMagrabi2019ArtificialIntelligenceClinical,CorpusRichardson2019BuildingMaintainingTrust,FriedmanFlynn2019Computable,CorpusWasylewicz2018ClinicalDecisionSupport,LehmannDowns2018Desiderata,AdlerMilsteinEtAl2018ManagingCBK,FlynnEtAl2018KORO,Wilkinson2016FAIR,MoreauEtAl2015PROV,CorpusPhansalkar2013CriteriaAssessingHigh,Ciccarese2013PAV,CorpusZhou2012StudyDiverseClinical,CorpusHemens2011ComputerizedClinicalDecision,CorpusWright2011DevelopmentEvaluationComprehensive,CorpusMoxey2010ComputerizedClinicalDecision,CorpusKawamoto2009NationalClinicalDecision,CorpusSchnipper2008SmartFormsElectronic,CorpusBodenreider2008BiomedicalOntologiesAction,CorpusSittig2006SurveyFactorsAffecting,Peleg2003GuidelineModels,CorpusAchour2001UmlsBasedKnowledge,CorpusThomas1999EvaluationInternetBased}

\bibliography{bibliography/references,bibliography/literature-review-corpus}

\pagebreak
\end{document}